\documentclass[twocolumn]{aastex631}

\usepackage{graphicx}

\graphicspath{{./}{figures/}}

\usepackage{color}

\usepackage{soul}
\setstcolor{red}
\usepackage[normalem]{ulem}

\begin{document}

\title{From Starburst to Quenching: Physical Properties of Extremely Compact Starbursts at z$\sim$0.1}

\author{Can Xu}
\affiliation{School of Astronomy and Space Science, Nanjing University, Nanjing, Jiangsu 210093, China}
\affiliation{Key Laboratory of Modern Astronomy and Astrophysics, Nanjing University, Ministry of Education, Nanjing 210093, China}
%\email{canxu@smail.nju.edu.cn}

\author{Mengyuan Xiao}
\affiliation{Department of Astronomy, University of Geneva, Chemin Pegasi 51, 1290 Versoix, Switzerland}

\author{Min Bao}
\affiliation{School of Physics and Technology, Nanjing Normal University, Nanjing 210023, People's Republic of China}
\affiliation{School of Astronomy and Space Science, Nanjing University, Nanjing, Jiangsu 210093, China}
\affiliation{Key Laboratory of Modern Astronomy and Astrophysics, Nanjing University, Ministry of Education, Nanjing 210093, China}

\author{Qiusheng Gu}
\affiliation{School of Astronomy and Space Science, Nanjing University, Nanjing, Jiangsu 210093, China}
\affiliation{Key Laboratory of Modern Astronomy and Astrophysics, Nanjing University, Ministry of Education, Nanjing 210093, China}

\author{Longji Bing}
\affiliation{Astronomy Centre, University of Sussex, Falmer, Brighton BN1 9QH, UK}

\author{Yong Shi}
\affiliation{School of Astronomy and Space Science, Nanjing University, Nanjing, Jiangsu 210093, China}
\affiliation{Key Laboratory of Modern Astronomy and Astrophysics, Nanjing University, Ministry of Education, Nanjing 210093, China}

\author{Yifei Jin}
\affiliation{Westlake University, 600 Dunyu Road, Xihu District, Hangzhou, Zhejiang 310030 PR China}

\author{Shiying Lu}
\affiliation{School of Astronomy and Space Science, Nanjing University, Nanjing, Jiangsu 210093, China}
\affiliation{Key Laboratory of Modern Astronomy and Astrophysics, Nanjing University, Ministry of Education, Nanjing 210093, China}

\correspondingauthor{Mengyuan Xiao, Min Bao, Qiusheng Gu}
\email{mengyuan.xiao@unige.ch\;\\
       mbao@nnu.edu.cn\;\\
       qsgu@nju.edu.cn}

\begin{abstract}

The compaction phase plays a crucial role in galaxy evolution, as it is strongly linked to star formation activities and structural transformation. We have identified a sample of extremely compact starburst galaxies (eCSBs) at low redshift~(z$\sim$0.1), which represent this critical evolutionary stage. 
These eCSBs are massive outliers with intense star formation and high infrared luminosities comparable to (U)LIRGs, while their structure already resembles quiescent galaxies. To investigate their molecular gas properties, we conducted IRAM 30m observations of $^{12}$CO J = 1--0 and $^{12}$CO J = 2--1 emission lines. Our results indicate that eCSBs exhibit a notably low molecular gas fraction~($\sim3\%$), and short gas depletion time~($\sim$ 20 Myr), suggesting that these galaxies are rapidly exhausting their remaining gas reservoir. 
Compared to normal (U)LIRGs, eCSBs show systematically lower $^{12}$CO(2-1)/$^{12}$CO(1-0) ratio~($R_{21} \sim 0.65 \pm 0.06$), similar to main sequence galaxies. 
The relatively low CO excitation may be associated with their high central stellar mass densities.
These findings provide new insight into the molecular gas properties of galaxies during the compaction phase, highlighting their unique condition and rapid evolution toward quiescence.

\end{abstract}


\section{Introduction} 
\label{sec:intro}

Understanding the physical process of galaxy evolution remains one of the major challenges in extragalactic astronomy. 
The Hubble morphological sequence~\citep{hubble1926} originally classifies galaxies according to the structure, ranging from ellipticals to spirals. But subsequent studies have shown that this structural distinction is tightly connected to other physical properties, such as color and star formation activity. These correlations give rise to the well-known bimodality between the blue, star-forming galaxies (SFGs) and the red, quiescent galaxies (QGs)~\citep{Strateva2001AJ....122.1861S,kauffmann2003MNRAS.341...33K,Baldry2004ApJ...600..681B}.
Morphologically, SFGs are typically disk-like and more extended than the compact, spheroid-dominated QGs~\citep{wuyts2011}.

Although the connection between galaxy morphology and star formation has become well established through large surveys such as the Sloan Digital Sky Survey (SDSS), the physical processes linking structural transformation to galaxy quenching remain incompletely understood.
It is widely accepted that the dense central regions of galaxies are strongly correlated with galaxy quenching, typically driven by dissipative processes such as major mergers~\citep{hopkins2008} and gravitational instabilities~\citep{dekel2009}.  
During dissipative processes, gas inflows towards the inner kiloparsec region, triggering intense star formation. The morphology of galaxies also undergoes dramatic change in a relatively short period due to gravitational interaction and the rapid growth of central stellar mass during starburst~\citep{hopkins2008}, which favors bulge formation. This phase is commonly referred to as ``compaction". Almost all massive galaxies undergo this process before quenching at high redshift~\citep{Tacchella2016,Zolotov2015}.

It is generally assumed that the bulge component in SFGs differs from the disk component. Kinematical evidence of non-circular orbits suggests that bulges are more similar to elliptical galaxies. The central stellar surface density of galaxies, often referred to as the stellar surface density within the central 1 kpc ($\Sigma_{1\mathrm{kpc}}$; \citealt{cheung2012}), provides a reliable tracer of bulge growth~\citep{Fang2013,Barro2017}.
Increasing evidence has shown that galaxies with high $\Sigma_{1\mathrm{kpc}}$ are more likely to be quiescent~\citep{cheung2012,Barro2017,Tacchella2016,woo2019,suess2021}.
Therefore, the threshold of $\Sigma_{1\mathrm{kpc}}$ serves as an effective diagnostic to distinguish between star-forming and quiescent galaxies.

Cosmological simulations and analytic models have proposed a physical pathway for this evolution: gas-rich processes—such as violent disk instabilities, dissipative gas inflows, wet mergers, or stream-fed compaction, can drive gas toward the galactic center, triggering intense central star formation and a rapid increase in $\Sigma_{1\mathrm{kpc}}$. This “compaction” phase is subsequently followed by morphological stabilization of the disk, feedback from the central region (including AGN activity), and/or gas depletion, all of which contribute to the suppression of star formation and the emergence of a compact quiescent remnant~\citep{dekel2014MNRAS.438.1870D,Zolotov2015,Martig2009}.
Together, these observational and theoretical results suggest that increasing $\Sigma_{1\mathrm{kpc}}$ can precede and even drive the quenching of star formation—commonly referred to as the compaction–quenching sequence.
This, in turn, supports the existence of a distinct compaction phase in the evolutionary pathway of galaxies.

A significant population of compact QGs are identified at high redshift~($z \gtrsim 2$)~\citep{naab2007,oser2010,Barro2013, almaini2017MNRAS.472.1401A}. Their structural properties indicate that they have undergone compaction phase.~\citet{Barro2013} proposed two evolutionary pathways for galaxies: the first path ($z \gtrsim 2$), galaxies undergo a rapid compaction and quenching phase, forming compact QGs that subsequently grow in size during their quiescent evolution;
and the second path (z$<$2), more extended star-forming galaxies evolve directly into extended QGs without experiencing a compaction phase.
Thus, compact QGs are expected to have undergone a compaction phase and evolved from compact star-forming progenitors~\citep{ newman2012ApJ...746..162N,belli2015ApJ...799..206B}. In fact, compact SFGs have been observed at $z\sim2$--3~\citep{wuyts2011,whitaker2012, barro2014,Spilker2016,Kaviraj2013}. The number density of these galaxies appears to decline toward lower redshift~\citep{poggianti2013ApJ...777..125P}, suggesting that the dissipative process responsible for rapid central mass build-up were more prevalent at high-redshift~\citep{ lu2019RAA....19..150L}. Understanding whether remnants of such early compaction events persist in the nearby Universe therefore requires studies of compact SFGs at low redshift.

In the nearby universe, only a few studies have focused on compact SFGs~\citep{Tacchella2016,Jaskot2013ApJ...766...91J,Izotov2011ApJ...728..161I}.
\citet{lapiner2023} investigated these low-redshift systems and found that their formation pathways differ from those of their high-redshift ($z\gtrsim2$) counterparts, which are often linked to smooth cold-stream accretion. They further suggested that at $z<2$, the rapid gaseous compaction phase is more likely to be triggered by violent, event-driven mechanisms, such as gas-rich mergers or violent disk instabilities.
However, despite these recent advances, the molecular gas properties of galaxies undergoing compaction remain poorly constrained. Direct measurements of their molecular gas reservoirs and excitation are still scarce, limiting our understanding of how molecular gas evolves immediately prior to quenching.

In this study, we identify a class of extremely compact starburst galaxies (eCSBs) at z$\leq$0.15. These are massive galaxies (M$_* > 10^{11}\;\mathrm{M}_{\odot}$) with enhanced star formation rates ($\geq 3\sigma$ above the main sequence), and are characterized by high central stellar mass densities ($\Sigma_{1\mathrm{kpc}}$), indicating strongly compact morphology (see Section 2 for details).
Investigating compact SFGs in the low-redshift universe is crucial, as it not only illuminates the formation pathways of massive compact QGs at high redshift but also sheds light on the ongoing evolutionary pathways of galaxies in the present epoch.
We present observations using the IRAM 30m telescope to study CO(1-0) and CO(2-1) emission lines.

The layout of this paper is as follows: In Section 2, we provide a detailed description of the sample selection process. In Section 3, we outline the observational setup and present the data reduction procedure. 
In Section 4, we present our main results, including the $L'_{\rm CO}$–$L_{\rm IR}$ relation, molecular gas fractions, depletion times, and the distribution of CO line ratios ($R_{21}$) across different galaxy properties.
Finally, Section 5 provides a summary of our findings. Throughout this work, we adopt a cosmology with $\Omega_{\rm m}= 0.3$, $\Omega_{\Lambda}= 0.7$ and $\rm H_{0}  = 70 km s^{-1}  Mpc^{-1} $. Magnitudes are provided in the AB system \citep{Gunn1972}.

%%%%%%%%%%%%%%%%%%%%
\begin{figure}[ht]
\centering
\includegraphics[scale=0.6
]{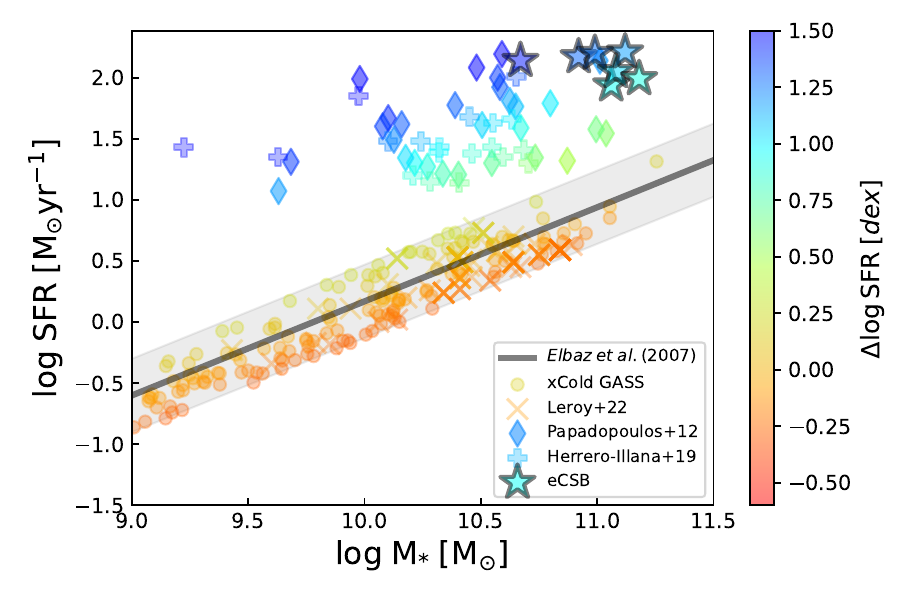}
\caption{
The star formation rate (SFR) versus stellar mass ($M_*$) diagram for the eCSBs and the literature reference comparison samples. 
Our eCSB sample is shown as red stars. 
Comparison samples include MS galaxies from xCold GASS (orange circles; \citealp{Saintonge2017}) and \citet{leroy2022} (brown crosses), along with (U)LIRGs from \citet{papa2012} (light blue plus signs)  and \citet{herrero2019} (gray triangles). 
The solid gray line indicates the local MS relation from \citet{elbaz2007A&A...468...33E}, with the shaded region marking its scatter. All data points are color-coded by their offset from the star-forming main sequence ($\Delta \log \mathrm{SFR}$), as indicated by the colorbar.
}
\label{SFR-M}
\end{figure}
%%%%%%%%%%%%%%%%%%%%

%%%%%%%%%%%%%%%%%%%%
\begin{figure}[htp!]
\centering
\includegraphics[scale=0.2]{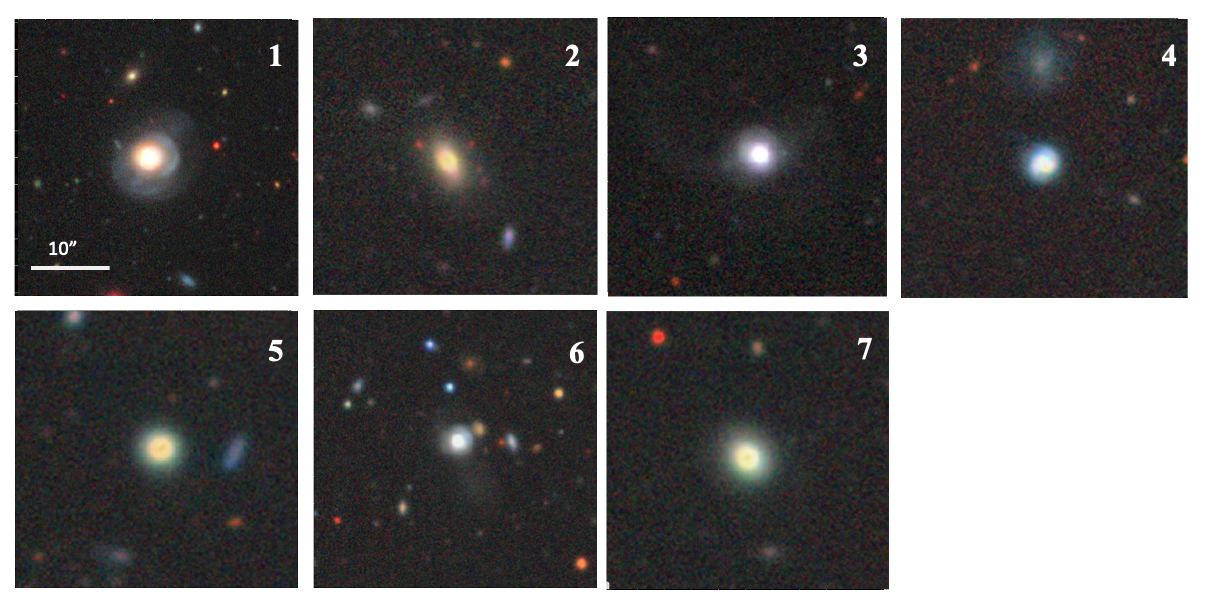}
\caption{RGB composite images of the seven eCSBs, with red, green, and blue corresponding to the \(r\)-, \(g\)-, and \(z\)-bands, respectively. These images are based on the DESI Legacy Imaging Surveys \citep{desi2019}.}
\label{DESI}
\end{figure}
%%%%%%%%%%%%%%%%%%%%

%%%%%%%%%%%%%%%%%%%%%%%%%%%%%%%%%%%%%%%%%%%%%%%%
\begin{table*}[htp]
\caption{{\bf Physical parameters of 7 extremely compact starburst galaxies}}\label{Tab}
\centering
\setlength{\tabcolsep}{1.2mm}  
\renewcommand{\arraystretch}{1.0} 
%---------------------------------------------------------------
\begin{tabular}{c|c|c|c|c|c|c|c|c|c|c}
\hline\hline 
ID & Name & RA & DEC & Redshift & $\log M_*$ & SFR & $\log L_{\mathrm{IR}}$ & $M_{\mathrm{gas}}$ & $t_{\mathrm{dep}}$ & $R_{21}$\\
   &  & [J2000] & [J2000] &  & [$M_\odot$] & [$M_\odot\,\mathrm{yr^{-1}}$] & [$L_\odot$] & [$10^{9}\,M_\odot$] & [Myr] &  \\
(1) & (2) & (3) & (4) & (5) & (6) & (7) & (8) & (9) & (10) & (11)\\
\hline
1 & IRAS 01249$-$0848  & 01:27:29 & $-$08:33:00 & 0.049 & 11.02$\pm$0.05 & 87   & 11.77 & 4.49$\pm$0.24 & 51.6$\pm$2.7 &  0.55$\pm$0.02\\
2 & IRAS F12028+2300 & 12:05:26 & +22:43:48 & 0.076 & 10.75$\pm$0.05 & 112  & 11.88 & 1.22$\pm$0.08 & 10.9$\pm$0.8 & 0.70$\pm$0.05 \\
3 & IRAS 13167+1336  & 13:19:14 & +13:20:24 & 0.096 & 10.53$\pm$0.04 & 138  & 11.97 & 1.56$\pm$0.15 & 11.3$\pm$1.1 & 0.80$\pm$0.06 \\
4 & IRAS 14575+3256  & 14:59:36 & +32:45:00 & 0.114 & 10.53$\pm$0.10 & 148  & 12.00 & 2.82$\pm$0.16 & 19.1$\pm$1.1 & 0.65$\pm$0.05 \\
5 & IRAS 08482+5247  & 08:51:50 & +52:36:00 & 0.132 & 10.98$\pm$0.07 & 162  & 12.04 & 4.89$\pm$0.26 & 30.2$\pm$1.6 & 0.72$\pm$0.04 \\
6 & IRAS 11347+2033  & 11:37:23 & +20:16:48 & 0.135 & 11.00$\pm$0.10 & 158  & 12.03 & 3.76$\pm$0.32 & 23.8$\pm$2.0 & 0.49$\pm$0.05 \\
7 & IRAS 14370+6254  & 14:38:13 & +62:42:00 & 0.115 & 10.98$\pm$0.04 & 98   & 11.82 & 3.18$\pm$0.34 & 32.4$\pm$3.5 & 0.67$\pm$0.04\\
\hline
\end{tabular}

\vspace{2mm}
\begin{minipage}{0.96\textwidth}
\small
\textbf{Notes.} 
(1) ID in the eCSB sample; 
(2) galaxy name; 
(3–4) right ascension and declination [J2000]; 
(5–8) redshift, stellar mass, SFR, and infrared luminosity $L_{\mathrm{IR}}$ are adopted from the MPA–JHU Value-Added Catalog~\citep{brinchmann2004MNRAS.351.1151B,kauffmann2003MNRAS.341...33K} and the Imperial IRAS–FSC Redshift Catalog~\citep{wangrowan2009}; 
(9) molecular gas mass $M_{\mathrm{gas}}$ is derived from the CO(1–0) luminosity assuming a standard conversion factor $\alpha_{\mathrm{CO}} = 0.8~M_\odot~(\mathrm{K\,km\,s^{-1}\,pc^2})^{-1}$; 
(10) gas depletion timescale $t_{\mathrm{dep}} = M_{\mathrm{gas}} / \mathrm{SFR}$; 
(11) CO line ratio $R_{21} = L'_{\mathrm{CO(2-1)}} / L'_{\mathrm{CO(1-0)}}$.

\end{minipage}
\end{table*}
%%%%%%%%%%%%%%%%%%%%%%%%%%%%%%%%%%%%%%%%%%%%%%%%

%%%%%%%%%%%%%%%%%%%%%%%%%%%%%%%%%%%%%%%%%%%%%%%%
\begin{table*}[htp]
\caption{{\bf \textbf{CO(1–0) and CO(2–1) Line Measurements for the IRAM 30m eCSB Sample}}  }
\label{CO-tab}
\centering
\setlength{\tabcolsep}{3.5mm}{
\begin{tabular}{c|cccc|cccc}
\hline\hline
ID & $\rm I_{CO(1-0)}$ & RMS & S/N & $\Delta V_{\rm CO(1-0)}$ 
   & $\rm I_{CO(2-1)}$ & RMS & S/N & $\Delta V_{\rm CO(2-1)}$ \\

 & [$\rm K\,km\,s^{-1}$] & [mK] & & [$\rm km\,s^{-1}$] 
 & [$\rm K\,km\,s^{-1}$] & [mK] & & [$\rm km\,s^{-1}$] \\

(1) & (2) & (3) & (4) & (5) & (6) & (7) & (8) & (9) \\
\hline
1 & 14.8$\pm$0.8  & 3.16 & 16 & 447 & 33.1$\pm$1.0 & 5.12 & 25 & 427 \\
2 & 1.65$\pm$0.11 & 1.04 & 4  & 562 & 5.0$\pm$0.4  & 2.16 & 7  & 658 \\
3 & 1.31$\pm$0.13 & 1.45 & 4  & 482 & 5.21$\pm$0.32 & 4.01 & 5  & 513 \\
4 & 1.67$\pm$0.09 & 1.38 & 7  & 370 & 4.56$\pm$0.15 & 1.81 & 12 & 423 \\
5 & 2.14$\pm$0.11 & 1.09 & 6  & 526 & 6.01$\pm$0.10 & 2.11 & 15 & 535 \\
6 & 1.57$\pm$0.14 & 0.99 & 3  & 583 & 2.87$\pm$0.22 & 2.51 & 4  & 474 \\
7 & 1.84$\pm$0.20 & 1.19 & 6  & 506 & 4.5$\pm$0.4   & 1.80 & 10 & 510 \\
\hline
\end{tabular}}
\vspace{2mm}
\begin{flushleft}
\textbf{Notes.} 
(1) Source ID in the eCSB sample. 
(2) and (6) Velocity-integrated line intensities and their statistical uncertainties for CO(1–0) and CO(2–1), respectively. 
(3) and (7) Root-mean-square (RMS) noise levels measured at a spectral resolution of $\Delta V = 20$ km\,s$^{-1}$. 
(4) and (8) Signal-to-noise ratios (S/N) for CO(1–0) and CO(2–1) detections. 
(5) and (9) FWHM line widths derived from Gaussian fitting for CO(1–0) and CO(2–1), respectively.
\end{flushleft}
\end{table*}
%%%%%%%%%%%%%%%%%%%%%%%%%%%%%%%%%%%%%%%%%%%%%%%%

%%%%%%%%%%%%%%%%%%%%
\begin{figure*}[htp!]
\centering
\includegraphics[scale=0.3]{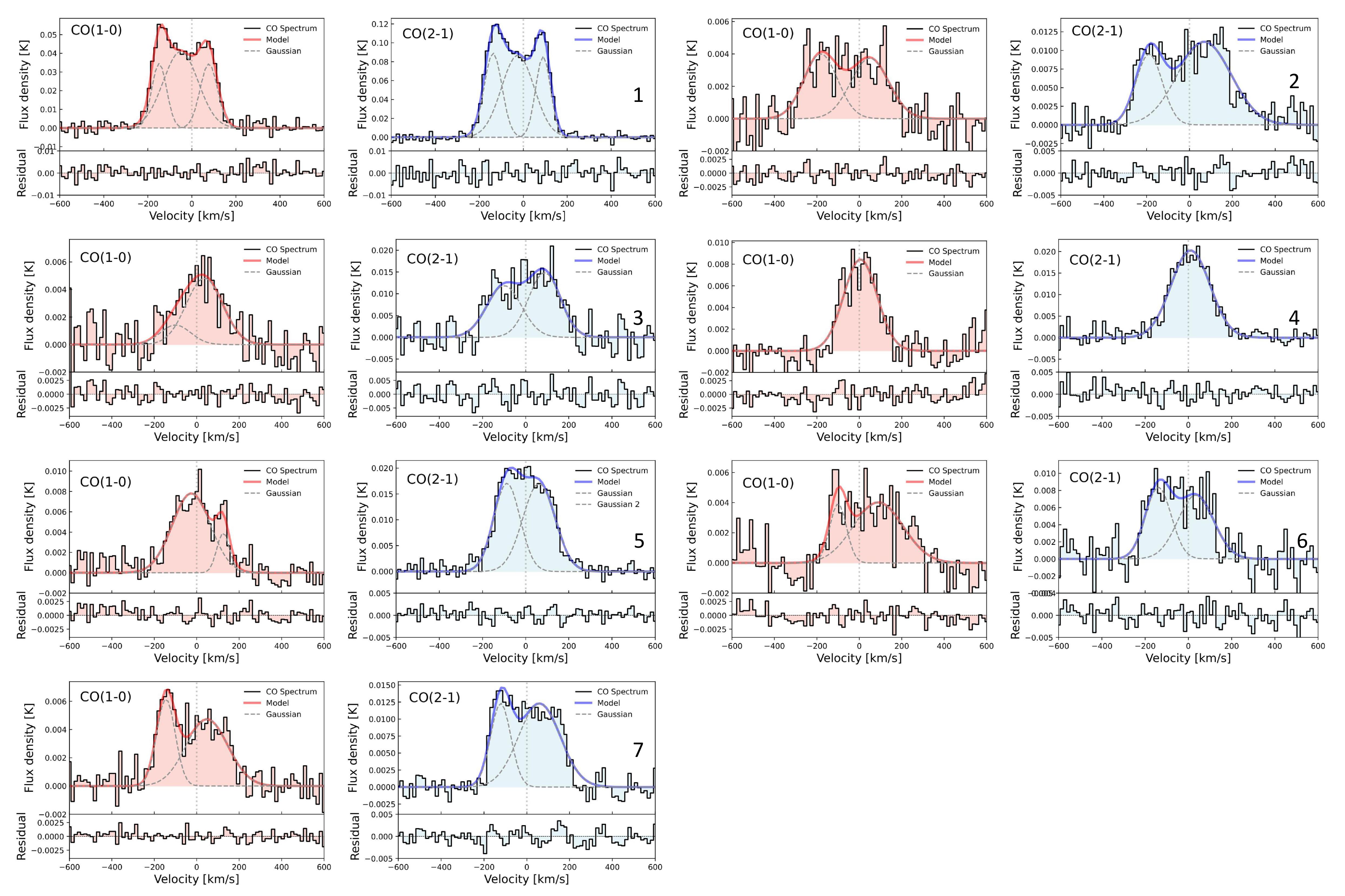}
\caption{Integrated, continuum-subtracted CO(1–0), CO(2–1) spectra, including only the galaxies observed by IRAM 30m. We show in red the CO(1–0) transition, in blue the CO(2–1) transition.}
\label{CO}
\end{figure*}
%%%%%%%%%%%%%%%%%%%%

\section{Sample selection} \
Our sample of massive, extremely compact starburst galaxies is selected from Sloan Digital Sky Survey, Data Release 8~(DR8) MPA-JHU Value-Added Catalog~\citep{brinchmann2004MNRAS.351.1151B,kauffmann2003MNRAS.341...33K}, from which we retrieve the redshift, stellar mass and star formation rate (hereafter SFR). 
Firstly, we select galaxies with stellar masses M$*>10^{10.5}\;M{\odot}$ and star formation rates exceeding $3\sigma$ above the star-forming main-sequence relation~\citep{elbaz2007A&A...468...33E} (Figure~\ref{SFR-M}), resulting in 2391 massive star-forming galaxies drawn from an initial catalog of $\sim$1.8 million sources.
Secondly, we cross-match this sample with the NASA–Sloan Atlas catalog \citep{blanton2011} and obtain structural parameters, Sérsic index ($n$) and the r-band effective radius ($R{\mathrm{e}}$) under single S{\'e}rsic model, for 261 of the 2391 galaxies. Thus we can estimate central surface density $\Sigma_{1kpc}$. $\Sigma_{1kpc}$ is defined as\[
\Sigma_{1 \mathrm{kpc}}=\frac{M_{*} \gamma\left(2 n, b_{n} R_{\mathrm{e}}^{-1 / n}\right)}{\pi},
\]
where $M_{*}$ is stellar mass, $b_{n}$ can be estimated as $b_{n}$=1.9992n-0.3271~\citep{graham2005}. $\gamma(2 n, b_{n} R_{\mathrm{e}}^{-1 / n})$ is the incomplete gamma function. Global compactness parameter $\Sigma_{1.5}$ is derived as $M_*/R_e^{1.5}$. We select galaxies based on the criteria of $\log\Sigma_{1\mathrm{kpc}} > 9.5~M_{\odot}~\mathrm{kpc}^{-2}$ and $\log\Sigma_{1.5} > 10.3~M_{\odot}~\mathrm{kpc}^{-1.5}$ \citep{Wisnioski2018}.
Finally, seven galaxies are chosen, all of which are found to lie at $z\leq0.15$.
The RGB composite (g–r–z) images of our sample, derived from the DESI Legacy Imaging Surveys~\citep{desi2019}, are presented in Figure~\ref{DESI}. 
The total infrared luminosities of these galaxies, as estimated from the Imperial IRAS-FSC Redshift Catalog~\citep{wangrowan2009}, exceed $10^{11.7}\;L_{\odot}$, thereby placing them in the class of (ultra-)luminous infrared galaxies (hereafter (U)LIRGs).

In the following section, we compare the physical properties of our eCSBs with those of reference galaxies. To build a control sample, we compile galaxies from previous CO surveys, including xCOLD GASS~\citep{Saintonge2017} and \citet{leroy2022}.
We identify main-sequence (MS) galaxies as those lying within $\pm0.3$ dex of the star-forming main sequence of~\citet{ elbaz2007A&A...468...33E}, based on their stellar mass.
We also incorporate a (U)LIRGs comparison sample from \citet{papa2012}, \citet{herrero2019}, and \citet{montoya2023}. These galaxies have high infrared luminosities ($L_{\mathrm{IR}}>10^{11-12}~L_\odot$), and we generally select those that lying above the MS relation by more than 0.3 dex.

Figure~\ref{SFR-M} shows the locations of our eCSB sample and the comparison galaxies relative to the star-forming MS relation on the SFR--M$_*$ diagram.
This selection provides a consistent framework for comparing the eCSBs with well-studied local galaxy populations. Throughout this work, the adopted MS and (U)LIRG samples are used as reference comparison populations for interpreting the molecular gas properties of the eCSBs.

\section{Data}
\subsection{Observation}

We observed seven galaxies with the IRAM 30,m telescope during the 2019B semester (Project ID 204-19; PI: Mengyuan Xiao). The observations were carried out with the Eight Mixer Receiver (EMIR) in wobbler-switching (WSW) mode (EMIR E90/E230; \citealt{CARTER2012}), which simultaneously covers the $^{12}$CO($J=1-0$) and $^{12}$CO($J=2-1$) lines. We employed the fast Fourier transform spectrometer (FTS) backend, which provides a frequency resolution of 200 kHz. The half-power beam width (HPBW) is $\sim21.4''$ at 115,GHz ($^{12}$CO $J=1-0$) and $\sim10.7''$ at 230 GHz ($^{12}$CO $J=2-1$), respectively.

Each galaxy was observed at a single pointing centered on its optical coordinates, with an on-source integration time of $\sim$2–3 hr per target. The achieved RMS noise levels and the corresponding signal-to-noise ratios for each spectrum are listed in Table~\ref{CO-tab}.

\subsection{Data Reduction} \label{datareduction}

We reduced the IRAM 30m data using the Continuum and Line Analysis Single-dish software (CLASS) package. The spectra were smoothed to a velocity resolution of 20~km~s$^{-1}$ for the $^{12}$CO J=1--0 and J=2--1 lines, which is sufficient to characterize their line profiles. Aperture corrections were not applied, as all galaxies have optical diameters smaller than the IRAM 30m beam at 230~GHz, making any such correction negligible.
The primary beam efficiencies were set to the recommended values from the IRAM documentation: B$_{\mathrm{eff}} \sim 78\%$ at 115~GHz and 59\% at 230~GHz. The forward efficiencies are F$_{\mathrm{eff}}=0.95$ for $^{12}$CO J=1--0 and $0.91$ for $^{12}$CO J=2--1. The main-beam temperature is given by T$_{\rm mb}$ = (F$_{\mathrm{eff}}$/ B$_{\mathrm{eff}}$) $\times$ T$_A^*$.

To analyze the spectra, a first-order (linear) polynomial was used to fit and subtract the continuum. The emission-line regions, typically spanning a velocity range centered on the systemic velocity of each galaxy, were then fitted with either single- or multi-Gaussian models, depending on the residuals. This procedure allows us to derive key spectral properties for further analysis. The CO line intensities were calculated by integrating the fitted Gaussian profiles, and the results are shown in Fig~\ref{CO}.

Following \citet{solomon1997}, we estimate the CO line luminosity, $L'_{\rm CO}$ (in K km s$^{-1}$ pc$^2$), from the Gaussian fitting results using: 
\[
L'_{\mathrm{CO}} \; [\mathrm{K \; km \; s^{-1} \; pc^2}] = 3.25 \times 10^7 \times S_{\mathrm{CO,tot}}  v_{\mathrm{rest}}^{-2}  D^2  (1 + z)^{-1},
\]
where $S_{\mathrm{CO,tot}}$ is the total CO line flux (in Jy km s$^{-1}$). The measured main-beam brightness temperature ($T_{\rm mb}$) are converted to flux density using the standard IRAM 30m telescope conversion factor of 5 Jy K$^{-1}$ at 110~GHz. $D$ is the luminosity distance in Mpc, $z$ is the redshift, and $v_{\mathrm{rest}}$ is the rest frequency of the line in GHz.

The molecular gas mass ($M_{\mathrm{mol}}$) is derived from the CO luminosity $L'_{\mathrm{CO(1-0)}}$~\citep{solomon1997, bolatto2013}:
\[
M_{\mathrm{mol}} [M_{\odot}] = 1.36 \times \alpha_{\mathrm{CO}} \times L'_{\mathrm{CO}},
\] 
where the factor 1.36 accounts for helium. 

We adopt a constant CO-to-H$_{2}$ conversion factor, $\alpha_{\mathrm{CO}} = 0.8\;M_{\odot}\;(\mathrm{K\;km\;s^{-1}\;pc^2})^{-1}$ to estimate the molecular gas masses of our eCSB sample and the (U)LIRGs comparison galaxies. Although our eCSB show high SFR and L$_{IR}$ typical of (U)LIRGs (Table~\ref{Tab}), our primary justification for this "ULIRG-like" $\alpha_{\text{CO}}$ comes from their far-infrared colors. We examined the IRAS F(60 $\mu$m)/F(100 $\mu$m) flux density ratio, a proxy for dust temperature. The sample has a median ratio of $0.6$, placing it firmly in the "warm" regime (F(60)/F(100)$>$ 0.5), characteristic of dense starbursts such as (U)LIRGs. 
For the MS sample~\citep{Saintonge2017,leroy2022}, we adopt the published gas masses, which were derived using the Galactic CO-to-H$_2$ conversion factor, $\alpha_{\mathrm{CO,Gal}} \approx 4.3~M_\odot~(\mathrm{K~km~s^{-1}~pc^2})^{-1}$.
The line ratio between $^{12}$CO J=1--0 and $^{12}$CO J=2--1 is defined as $R_{21} = \frac{L'_{\mathrm{CO(2-1)}}}{L'_{\mathrm{CO(1-0)}}}$, and is calculated directly from the line intensities derived as described above.

%%%%%%%%%%%%%%%%%%%%
\begin{figure}[ht]
\centering
\includegraphics[scale=0.62]{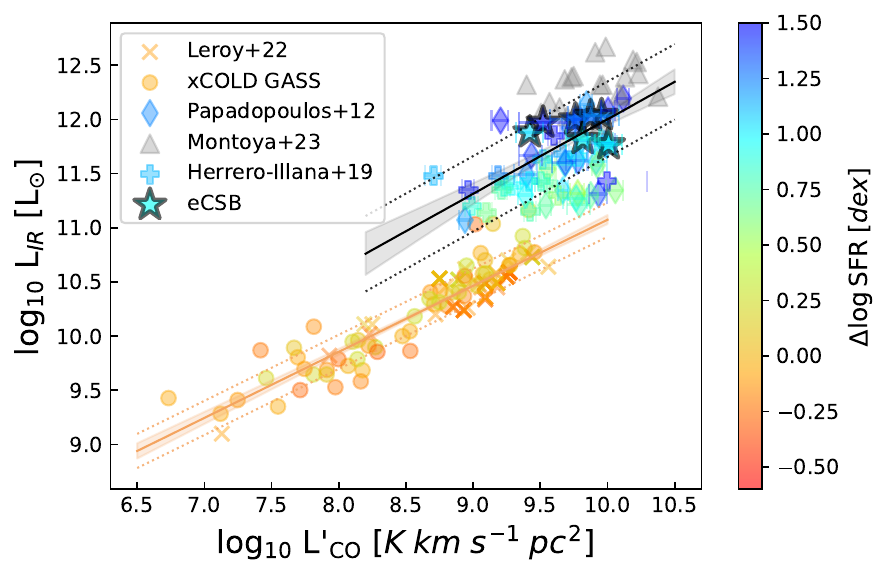}
\caption{The relationship between the infrared luminosity ($L_{\mathrm{IR}}$) and the CO(1-0) luminosity ($L'_{\mathrm{CO}}$). Our eCSB sample is indicated by stars. Comparison sample: MS galaxies from the xCOLD GASS survey (circles; \citealt{Saintonge2017}), \citet{leroy2022} (crosses); (U)LIRGs from \citet{papa2012} (diamonds), \citet{montoya2023} (triangles), and \citet{herrero2019} (pluses). All data points are color-coded by their offset from the star-forming main sequence ($\Delta \log \mathrm{SFR}$), as indicated by the colorbar. The solid orange and black lines represent the MCMC best-fit relations for the MS and (U)LIRG samples, respectively, with the corresponding dotted lines showing the 1$\sigma$ scatter for each fit.}
\label{CO-IR}
\end{figure}
%%%%%%%%%%%%%%%%%%%%

%%%%%%%%%%%%%%%%%%%%
\begin{figure}[ht]
\centering
\includegraphics[scale=0.65]{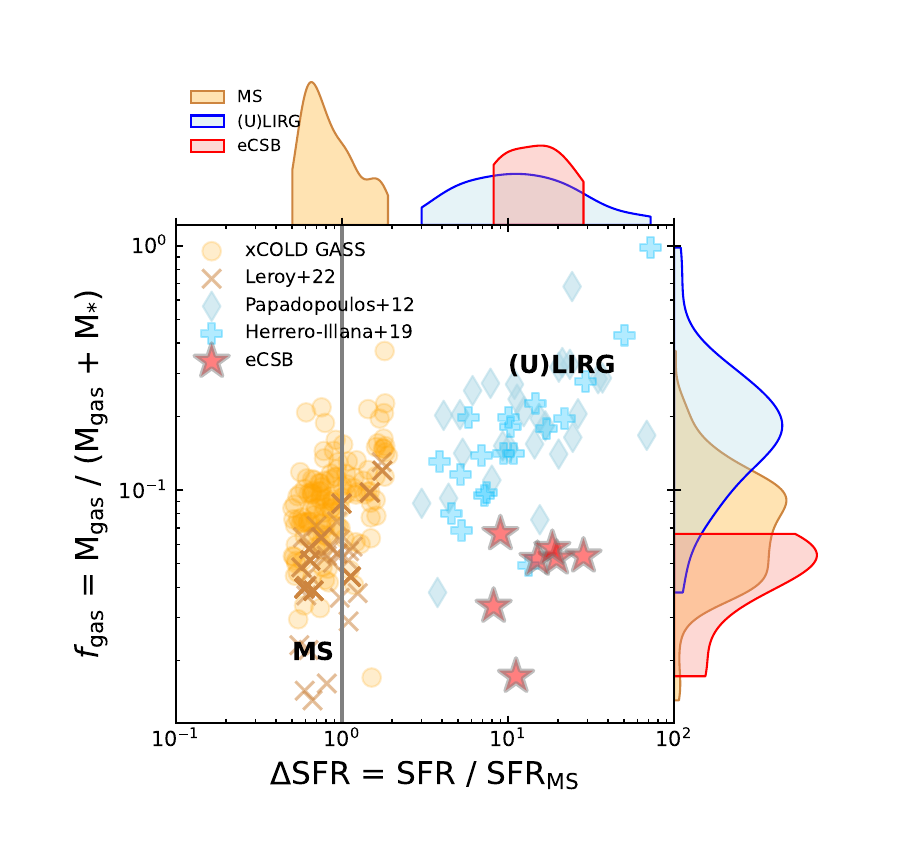}
\caption{The molecular gas fraction ($f_{\mathrm{gas}} = M_{\mathrm{gas}} / (M_{\mathrm{gas}} + M_*)$) is plotted as a function of the offset from the star-forming main sequence ($\Delta \mathrm{SFR} = \mathrm{SFR} / \mathrm{SFR}{\mathrm{MS}}$).
Our eCSB sample is shown as red stars. For comparison, we include main-sequence galaxies from the xCOLD GASS survey (orange circles; \citealt{Saintonge2017}) and \citet{leroy2022} (peru crosses), as well as (U)LIRGs from \citet{papa2012} (blue diamonds) and \citet{herrero2019} (cyan plus signs).
The vertical gray line marks the location of the main sequence ($\Delta \mathrm{SFR} = 1$).
The top and right subpanels show the normalized distributions (Kernel Density Estimates) of $\Delta \mathrm{SFR}$ and $f_{\mathrm{gas}}$ for the eCSB (red), MS (gold), and (U)LIRG (blue) samples, respectively.
The $\alpha_{\mathrm{CO}}$ conversion factors adopted to derive $M_{\mathrm{gas}}$ for each sample are described in Sect.~\ref{datareduction}.
}
\label{gasfraction}
\end{figure}
%%%%%%%%%%%%%%%%%%%%

\section{Results and Discussion }

\subsection{$L'_{\mathrm{CO}}$ versus $L_{\mathrm{IR}}$}

Before investigating molecular line ratios and their dependence on galaxy properties, we first explore the trends of the line luminosities that are used to compute these ratios. In Figure~\ref{CO-IR}, we compare the CO and IR luminosities of the eCSBs with those of (U)LIRGs~\citep{herrero2019,papa2012,montoya2023} and MS galaxies~\citep{Saintonge2017,leroy2022}. The observables L$'_{\mathrm{CO}}$ and L$_{\mathrm{IR}}$ serve as key indicators for estimating the molecular gas content and SFR of galaxies.

The L$'_{\mathrm{CO}}$ vs. L$_{\mathrm{IR}}$ relation is shown in Figure~\ref{CO-IR}. 
The eCSBs exhibit good agreement with the (U)LIRG samples from~\citet{herrero2019},~\citet{papa2012} and~\citet{montoya2023}.
We performed a correlation analysis on the combined eCSB and literature (U)LIRG samples and found a strong, statistically significant correlation~(Spearman coefficient $\rho = 0.82$, p-value of $1.8 \times 10^{-3}$).
The strong correlation indicates that the eCSBs lie close to the locus defined by the literature (U)LIRG sample in the $L'_{\rm CO}$--$L_{\rm IR}$ plane. We therefore perform a joint fit to these samples in order to characterize their common locus in this parameter space, and obtaining a best-fit relation of $\log L_{IR} = (0.693 \pm 0.122) \log L'_{CO} + (5.071 \pm 1.180)$ (solid black line), with an intrinsic scatter of $\sigma_{y} = 0.3$ dex.

Notably, the (U)LIRG sequence lies systematically above the MS galaxies on the L$'{\mathrm{CO}}$ vs. L${\mathrm{IR}}$ plane. An MCMC fit to the MS galaxies (data from \citet{leroy2022} and \citealt{Saintonge2017}) yields a distinct relation of $\log L_{IR} = (0.611 \pm 0.030) \log L'_{CO} + (4.962 \pm 0.264)$ (solid orange line), with a significantly smaller intrinsic scatter of $\sigma_{y} = 0.16$ dex. 
The differences in slope and normalization reflect the locations of the adopted MS and (U)LIRG comparison samples in the $L'_{\rm CO}$--$L_{\rm IR}$ plane.

Several studies~\citep[e.g.,][]{Daddi2010,Genzel2010,bolatto2013,narayanan2008,sargent2014, gao2004ApJ...606..271G, garcia2012A&A...539A...8G} have attributed this offset to variations in molecular gas properties, including higher gas densities, elevated turbulence, and more compact morphologies, which enhance star formation efficiency (SFE). Indeed, galaxies on the starburst sequence convert molecular gas into stars far more efficiently than MS galaxies with comparable molecular gas content. 
More recent work, however, suggests that this apparent separation should not be interpreted as evidence for two intrinsically distinct star-formation modes. Instead, molecular gas properties appear to evolve more continuously across the star-forming galaxy population~\citep[e.g.,][]{sargent2014,tacconi2020ARA&A..58..157T}. Accordingly, the adopted MS and (U)LIRG samples are used here as empirical reference populations that provide context for interpreting the molecular gas properties of the eCSBs, rather than as evidence for two distinct star-formation modes.

%%%%%%%%%%%%%%%%%%%%
\begin{figure*}[ht]
\centering
\includegraphics[scale=0.6]{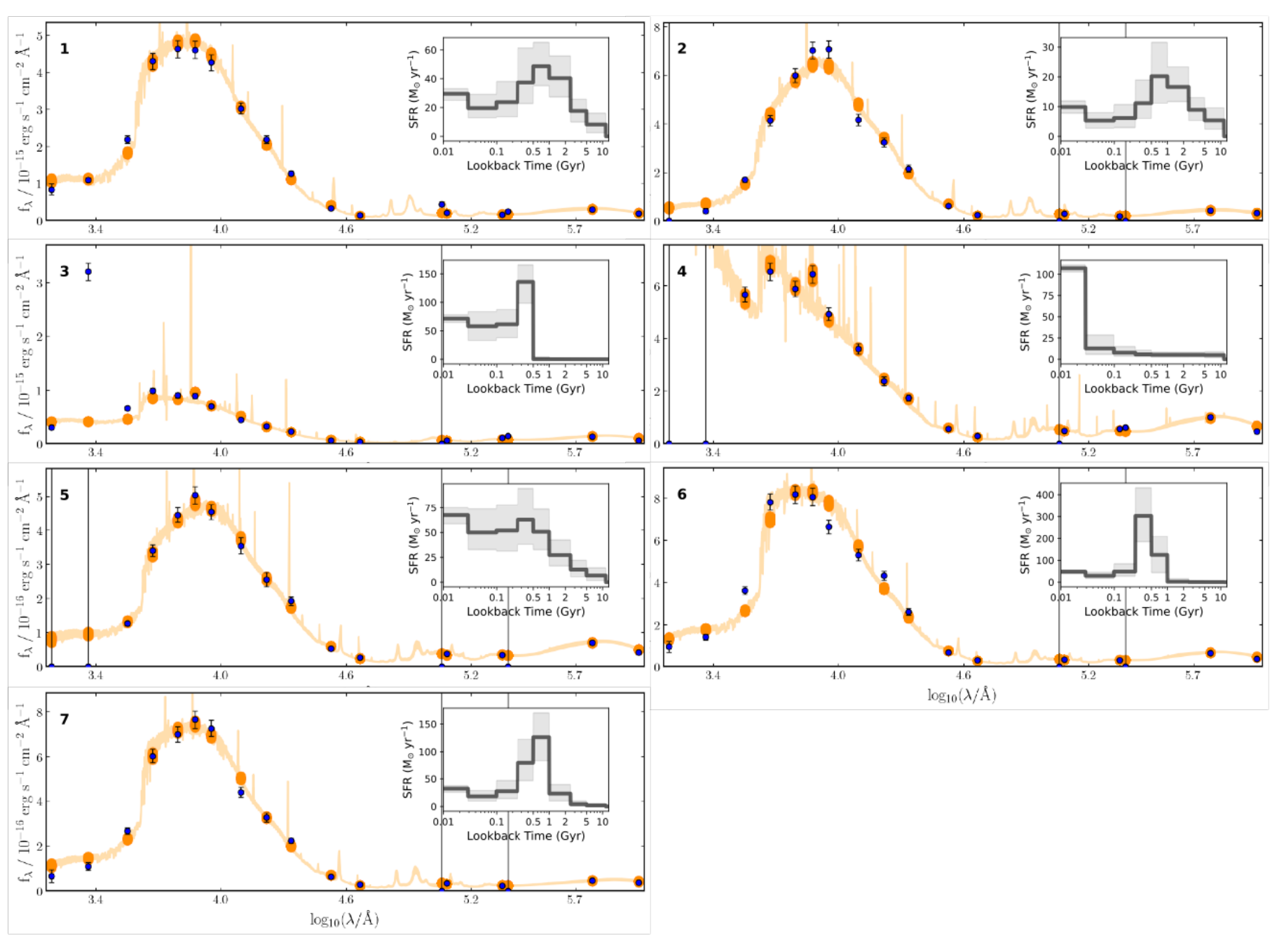}
\caption{Spectral energy distribution and star formation history of our eCSB sample.}
\label{sed1}
\end{figure*}
%%%%%%%%%%%%%%%%%%%%

\subsection{Gas Fraction}
We estimate the gas mass fractions as f$_{\mathrm{gas}} = \mathrm{M}_{\mathrm{gas}} / (\mathrm{M}_{\mathrm{gas}} + \mathrm{M}_*)$. Within our eCSB sample, f$_{\mathrm{gas}}$ has an average value of approximately 3\%~(see Figure~\ref{gasfraction}). An Anderson–Darling test yields a p-value of 0.91, indicating the gas fractions of eCSBs are statistically consistent with those of MS galaxies. Nevertheless, the star formation rates of our eCSBs are significantly higher, exceeding the MS by more than 3$\sigma$. This indicates that, despite comparable molecular gas content, eCSBs are experiencing much more intense star formation, likely driven by efficient gas consumption in their central regions.

When compared to (U)LIRGs from literature~(see Figure~\ref{gasfraction}), the p-value of 0.041 confirms a significant difference between the two sample, with statistical significance at the 5\% level. Our eCSBs exhibit significantly lower gas fractions than literature (U)LIRGs. Given that both the literature (U)LIRGs and eCSBs have comparable $L'_{\mathrm{CO}}$ (and thus $M_{\mathrm{gas}}$) and $L_{\mathrm{IR}}$ (and thus SFR; as shown in Figure~\ref{CO-IR}), the lower $f_{\mathrm{gas}}$ in eCSBs suggests that a significant fraction of their molecular gas has already been consumed prior to quenching.

We also estimate the gas depletion timescales, defined as $t_{\mathrm{dep}} = M_{\mathrm{mol}}/\mathrm{SFR}$~(Table~\ref{Tab}) and find that eCSBs exhibit considerably shorter gas depletion timescales compared to both MS galaxies and (U)LIRGs. The estimated $t_{\mathrm{dep,eCSB}}$ ranges between 10–60 Myr, whereas (U)LIRGs typically exhibit $t_{\mathrm{dep}}\sim$100 Myr~\citep{montoya2023}. And MS galaxies generally show $t_{\mathrm{dep,MS}} > 1~\mathrm{Gyr}$~\citep{leroy2022}. 

The short depletion time, together with their low gas fractions and elevated positions above the main sequence, indicate that eCSBs are undergoing intense, short-lived bursts of star formation. 
The rapid consumption of molecular gas is consistent with their elevated star formation activity, suggesting that these galaxies may evolve rapidly toward quiescence once their remaining gas reservoirs are exhausted.

This interpretation is further supported by the (rotation) dynamical timescales at the half-light radius ($t_{\mathrm{rot}} = 2\pi r_{e}/(v_{\mathrm{FWHM}}/2)$). Under the assumption that the molecular gas follows the stellar distribution, we estimate an upper limit for the dynamical timescale. For our eCSB sample, $t_{\mathrm{rot}}$ is estimated to range from 20–60 Myr. The fact that $t_{\mathrm{dep,eCSB}}$ is comparable to  $t_{\mathrm{rot}}$ confirms that these galaxies are undergoing rapid quenching within a single rotation period.

In addition, our spectral energy distribution (SED) fitting, based on multi-band photometry from the far-UV to the infrared and using non-parametric star-formation histories derived with BAGPIPES~\citep{feroz2008,feroz2009,feroz2013, carnall2018MNRAS.480.4379C}, provides independent evidence for recent intense star-formation activity. The reconstructed star formation histories~(SFHs; Fig.~\ref{sed1}) indicate that all objects experienced a pronounced burst within the past $\sim$1~Gyr. Together with the short molecular gas depletion timescales and low gas fractions, these results suggest that the remaining gas reservoirs have been largely consumed by their recent episodes of enhanced star formation.

%%%%%%%%%%%%%%%%%%%%
\begin{figure*}[!htp]
\centering
\includegraphics[scale=0.5]{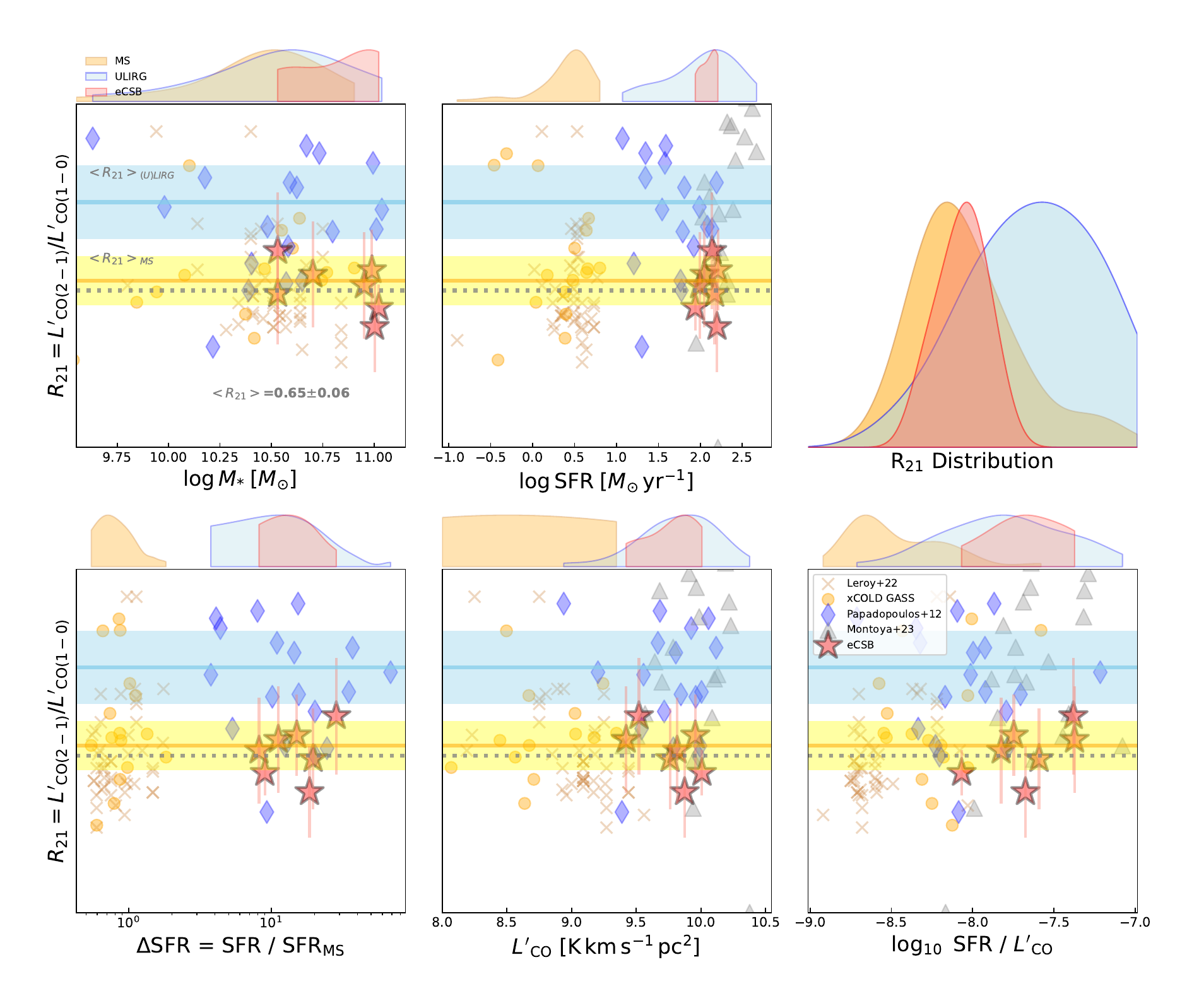}
\caption{ 
Scatter plots between the integrated $R_{21} \equiv L'_{\text{CO}(2-1)}/L'_{\text{CO}(1-0)}$ ratio and global galaxy properties. 
The six panels show $R_{21}$ as a function of (top row, left to right): stellar mass ($M_*$), star formation rate (SFR), and $R_{21}$ distribution of all sample; and (bottom row, left to right): SFR offset from the main sequence ($\Delta\text{SFR} = \text{SFR}/\text{SFR}_{\text{MS}}$), CO(1-0) luminosity ($L'_{\text{CO}}$), and star formation efficiency ($\text{SFR}/L'_{\text{CO}}$). 
The legend provides descriptions for the data points. Our eCSB sample is shown as red stars. 
The yellow and blue shaded regions represent the median $R_{21}$ value and Median Absolute Deviation~(MAD) for the MS and (U)LIRG sample, respectively. 
The black dotted line indicates the median value for our eCSB sample, $\langle R_{21} \rangle = 0.65 \pm 0.06$.
}

%\label{fig:R21_properties}
\label{R21-distribution}
\end{figure*}
%%%%%%%%%%%%%%%%%%%%

%%%%%%%%%%%%%%%%%%%%
\begin{figure}[!htp]
\centering
\includegraphics[scale=0.5]{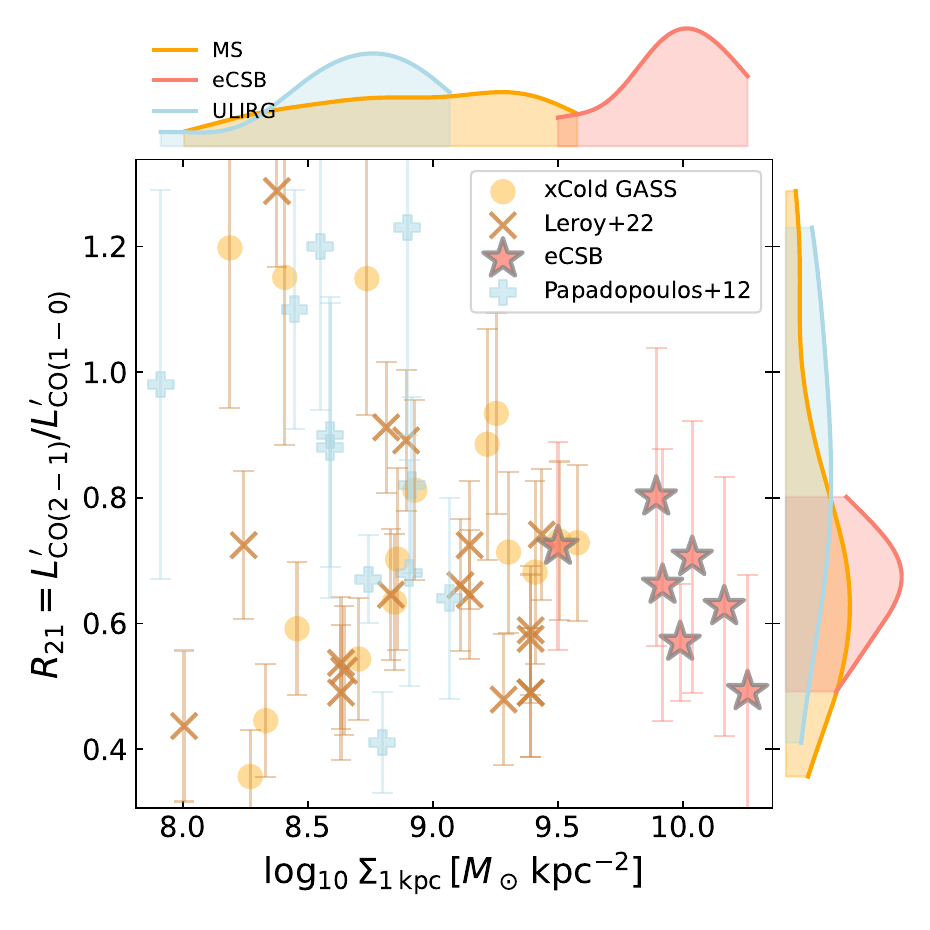}
\caption{Central 1kpc stellar mass surface density ($\log\Sigma_{1\text{kpc}}$) versus $R_{21}$. Red stars represent our eCSB sample; orange circles and brown crosses indicate main-sequence galaxies from~\citet{Saintonge2017} and \citet{leroy2022}, respectively; light blue plus signs correspond to (U)LIRGs from \citet{papa2012}. Top and right panels show the KDE distributions.
    }
\label{fig:density_comparison}
\end{figure}
%%%%%%%%%%%%%%%%%%%%

\subsection{The CO(2-1)/CO(1-0) Line Ratio ($R_{21}$)}

$R_{21}$, a widely adopted diagnostic of molecular‑gas excitation in galaxies, traces systematic variations in the bulk molecular interstellar medium (ISM) across different galaxy populations~\citep{keenan2024ApJ...975..150K, maeda2022ApJ...926...96M,israel2005A&A...438..855I}. Because the CO(2–1) transition originates from a higher rotational level than CO(1–0), variations in $R_{21}$ reflect changes in the excitation conditions of the molecular gas, indirectly tracing differences in gas density, excitation environment, or large-scale star-forming conditions~\citep{maeda2022ApJ...926...96M,denbrok2021,denbrok2022A&A...662A..89D}.
Surveys of star‑forming galaxies reveal that the luminosity‑weighted mean \(R_{21}\) is around \(0.6\)–\(0.7\), with modest enhancements (of order 10--20\%) observed in galaxy centers or regions of elevated star‐formation surface density~\citep{denbrok2021,denbrok2022A&A...662A..89D}.

In this section, we investigate how \(R_{21}\) correlates with both global galaxy properties—including stellar mass (\(M_*\)), star formation rate (SFR), offset from the star‑forming main sequence ($\Delta$SFR), CO luminosity (\(L'_{\rm CO}\)), and SFR/\(L'_{\rm CO}\)—and the central stellar mass surface density \(\Sigma_{1\,\rm kpc}\), which traces the compactness of the inner stellar structure. 
Since \(\Sigma_{1\,\rm kpc}\) is closely linked to bulge growth, compaction, and the depth of the central gravitational potential, it provides a physically motivated tracer of the pressure and density conditions that regulate molecular gas excitation~\citep{narayanan2014MNRAS.442.1411N}.
This combined approach allows us to examine how both large-scale galaxy evolution and central structural concentration influence the excitation state of molecular gas, as reflected in variations of \(R_{21}\).

\subsubsection{R$_{21}$ versus global parameters}

Figure~\ref{R21-distribution} presents the scatter plots of the galaxy-integrated CO line ratio (R$_{21}$) against various galaxy properties, following the approach outlined by~\citet{leroy2022}.

No significant trend is apparent in the first panel of Figure~\ref{R21-distribution}, likely due to the limited stellar mass distribution of MS galaxies and (U)LIRGs. However, in the remaining panels, a clear trend emerges.
In the comparison sample, (U)LIRGs and MS galaxies have broadly similar stellar-mass distributions (\emph{p}-value $\approx 0.1$), yet (U)LIRGs exhibit systematically higher SFR and $\Delta\text{SFR}$, larger molecular gas reservoirs, and shorter gas depletion times (SFR/$L'_{\mathrm{CO}}$) compared to the MS sample (\emph{p}-value $\sim 10^{-3}$)~\citep{rodighiero2011ApJ...739L..40R, elbaz2011A&A...533A.119E, solomon2005ARA&A..43..677S, saintonge2011MNRAS.415...32S,  tacconi2018ApJ...853..179T, Daddi2010, Genzel2010}. 
Given these global properties, the eCSBs are statistically consistent with the adopted (U)LIRG comparison sample (\emph{p}-value $\approx 0.25$).
For context, similar compact starburst systems have been studied at high redshift~\citealt{Barro2013,barro2014,Puglisi2019,tadaki2017ApJ...834..135T}).

As shown in Figure~\ref{R21-distribution}, the $R_{21}$ of (U)LIRGs are significantly higher than those of MS galaxies (\emph{p}-value $\sim 10^{-3}$), consistent with previous studies showing that elevated $R_{21}$ are associated with enhanced star-formation activity, higher star-formation efficiencies, and more excited molecular gas conditions \citep[e.g.,][]{yao2003ApJ...588..771Y, greve2014ApJ...794..142G, rosenberg2015ApJ...801...72R, kamenetzky2016ApJ...829...93K, lamperti2020ApJ...889..103L, denbrok2021, keenan2024ApJ...975..150K}. 
These results are consistent with the differences between the adopted MS and (U)LIRG comparison samples.
Given these observed difference, one might expect our eCSB sample to exhibit similarly elevated $R_{21}$ values.%\citep[e.g.,][]{condon1991ApJ...378...65C, greve2014ApJ...794..142G, diamond2021}. 

However, we find that the $R_{21}$ distribution of the eCSBs is statistically consistent with that of MS galaxies (\emph{p}-value~$\sim0.25$) and significantly lower than that of (U)LIRGs (\emph{p}-value~$\sim0.04$). 
This suggests that $R_{21}$ is not determined solely by global star-formation activity or the total molecular gas reservoir, but may also be sensitive to the local physical conditions within galaxies. 
In particular, variations in gas density, optical depth, radiation field, and the concentration of molecular gas may all influence CO excitation and therefore the observed $R_{21}$ ratio.
\citep[e.g.,][]{lamperti2020ApJ...889..103L,  narayanan2014MNRAS.442.1411N, leroy2022, denbrok2021}. 
To explore this possibility, we next focus on the central stellar mass surface density~($\Sigma_{1\,\mathrm{kpc}}$), an observational tracer of central compactness and the depth of the gravitational potential, and investigate its correlation with $R_{21}$.

\subsubsection{R$_{21}$ versus Density}

In the previous section, we found that the R$_{21}$ values of (U)LIRGs from the literature differ from those of our eCSBs.
We therefore investigate whether this difference is associated with the unusually high central stellar mass densities of the eCSBs.

Figure~\ref{fig:density_comparison} shows the distribution of R$_{21}$ and $\Sigma_{1\,\rm kpc}$ for the different galaxy populations. The MS and (U)LIRG samples exhibit largely overlapping $\Sigma_{1\,\rm kpc}$ distributions, with median values of $\log \Sigma_{1\,\rm kpc} \sim 8.875\;M_{\odot}\;kpc^{-2}$ (MAD = 0.413) and $\log \Sigma_{1\,\rm kpc} \sim 8.741\;M_{\odot}\;kpc^{-2}$ (MAD = 0.164), respectively. 
In contrast, the eCSBs are clearly offset toward much higher central densities, with a median of $\log \Sigma_{1\,\rm kpc} \approx 9.988\;M_{\odot}\;kpc^{-2}$ (MAD = 0.096), and are statistically distinct from both the MS and (U)LIRG populations (\emph{p}-values~$\sim10^{-3}$).

In the low‑ to moderate‑density regime, both the MS and (U)LIRG populations exhibit a large scatter in R$_{21}$, suggesting that a variety of excitation pathways and gas phases coexist in these environments. 
For instance, \citet{lamperti2020ApJ...889..103L} show that in low-redshift galaxies with lower gas densities and weaker radiation fields, the higher-$J$ CO transitions exhibit enhanced variation due to effects such as beam dilution, differences in cloud filling factors, and variations in FUV/H$_2$ heating. Similarly,~\citet{penaloza2017MNRAS.465.2277P} demonstrate using cloud-scale simulations that when densities are below the critical density for a given transition, excitation depends sensitively on local conditions—e.g., ambient radiation, turbulence, and optical depth—resulting in large intrinsic scatter in line ratios.

The eCSBs instead occupy a region of substantially higher central stellar mass density while exhibiting relatively low and narrowly distributed $R_{21}$ values.
This behavior is consistent with a scenario in which increased central compactness — and the deeper gravitational potential it implies — is associated with higher central gas pressures and optical depths.
Theoretical predictions and observational evidence regarding dense molecular environments both indicate that in high-density or high-pressure gas, $R_{21}$ can indeed be suppressed.
For example, \citet{papa2012} argue that in (U)LIRGs, where molecular gas may reside in very dense, optically thick, dynamically bound clouds, the CO ladders flatten and low-$J$ line ratios (including $R_{21}$) fall below the values expected from simple scaling relations.
Radiative transfer modeling by \citet{narayanan2014MNRAS.442.1411N} further shows that when gas densities exceed the critical densities of the transitions and optical depths become significant, higher-$J$ lines saturate or thermalize, thereby reducing their ratios relative to the ground state.
Likewise, \citet{Koda2012} find that in the dense regions of the spiral galaxy M51, the CO(3–2)/CO(1–0) ratio decreases in the highest surface-density zones, likely due to sub-thermal excitation or increased optical thickness.

Taken together, these results suggest that the relatively low $R_{21}$ values observed in eCSBs may reflect changes in the physical conditions of the molecular gas associated with strong central structural concentration.
However, because $\Sigma_{1,\rm kpc}$ traces the stellar distribution rather than the molecular gas directly, this interpretation remains suggestive rather than conclusive. 
Spatially resolved gas tracers or high-resolution CO excitation ladders are required to establish a direct causal link between central compactness and CO excitation conditions.

\section{Summary}

We have conducted a comprehensive analysis of high signal-to-noise $^{12}$CO J = 1–0 and J = 2–1 emission line spectra for eCSBs at z$\sim$0.1.
These systems likely represent a critical evolutionary phase in which structural compaction precedes the quenching of star formation.
We further compared the physical and molecular gas properties of eCSBs with those of MS galaxies and (U)LIRGs from the literature. The main findings of our study are summarized as follows:

\begin{enumerate}

\item \textbf{CO--IR luminosity relation:} 
eCSBs follow the well-established correlation between CO luminosity ($L'_{\rm CO}$) and infrared luminosity ($L_{\rm IR}$) defined by the adopted local (U)LIRG comparison sample.. 
In contrast, they are systematically offset from the selected local MS comparison sample.

\item \textbf{Gas fraction and depletion:} 
Despite having $L_{\rm IR}$ comparable to (U)LIRGs, eCSBs have lower total molecular gas fraction, approaching values typical of MS galaxies. SED-derived star-formation histories indicate that these galaxies have undergone a burst of star-forming activity within the past $\sim1$~Gyr. Combined with the short molecular gas depletion times ($\tau_{\rm dep} \sim 20$~Myr), these findings suggest that eCSBs are undergoing rapid gas consumption and are on a fast evolutionary path toward quenching.

\item \textbf{Suppressed CO excitation:} 
Although eCSBs resemble (U)LIRGs in global properties such as SFR, stellar mass, and SFR/$L'{\rm CO}$, their $R{21}$ values are systematically lower than those of (U)LIRGs and instead comparable to MS galaxies (eCSB vs.\ MS: $p\sim0.25$; eCSB vs.\ (U)LIRGs: $p\sim0.04$).

\end{enumerate}

The combination of low gas fraction, short gas depletion time, compact morphology, and suppressed $R_{21}$
identifies eCSBs as a critical transitional population along the compaction--quenching sequence, in which structural compaction precedes the quenching of star formation. The relatively low CO excitation observed in the eCSBs may reflect changes in the physical conditions of the molecular gas during this compact evolutionary stage.Our results suggest that, even among systems with broadly similar global properties, central structural concentration may be related to differences in molecular gas properties.

Future studies, particularly those employing larger and more representative galaxy samples together with spatially resolved molecular gas observations, will be essential for further clarifying the mechanisms that govern this rapid evolutionary stage.

\begin{acknowledgments}

This work is based on observations carried out under project number 204-19 with the IRAM 30m telescope. IRAM is supported by INSU/CNRS (France), MPG (Germany) and IGN (Spain). This work is suported by the National Natural Science Foundation of China~(No.12192222, 12192220 and 12121003), as well as by funding from the Swiss State Secretariat for Education, Research and Innovation (SERI) under contract number MB22.00072, as well as from the Swiss National Science Foundation (SNSF) through project grant 200020\_207349. 

\end{acknowledgments}

%\appendix

\bibliographystyle{aasjournal}
\bibliography{main.bib}

\end{document}